\documentclass[aps,nofootinbib]{revtex4}
\usepackage{amsmath,amssymb,amsbsy,amsfonts,latexsym,graphicx,hyperref}
\usepackage{color}
\usepackage{rotating}
\usepackage{graphicx,subfigure,color,array,slashed,wasysym}
\usepackage{caption}
\usepackage{dcolumn}% Align table columns on decimal point
\usepackage{bm}% bold math
\usepackage[mathlines]{lineno}% Enable numbering of text
\usepackage{hyperref}% add hypertext capabilities

\usepackage{bbold}

\newcommand{\be}{\begin{equation}}
\newcommand{\ee}{\end{equation}}
\newcommand{\ba}{\begin{array}}
\newcommand{\ea}{\end{array}}
\newcommand{\bea}{\begin{eqnarray}}
\newcommand{\eea}{\end{eqnarray}}
\usepackage{multirow}

\begin{document}

\begin{center}
{\bf The superconducting dome for holographic doped Mott insulator with hyperscaling violation}\\

\vspace{1.6cm}

Wenhe Cai $^{~a,b,*}$, Sang-Jin Sin $^{a,\dag}$\let\thefootnote\relax\footnotetext{* whcai@shu.edu.cn, \dag sjsin@hanyang.ac.kr}\\
\vspace{0.8cm}

$^a${\it Department of Physics, Hanyang University, Seoul 04763, Korea} \\
$^b${\it Department of Physics, Shanghai University, Shanghai 200444,  China}
\vspace{1.6cm}

\begin{abstract}
We reconsider the holographic model  featuring a superconducting dome on the temperature-doping  phase diagram with a modified view on the role of the two charges. The first type charge with density $\rho_{A}$  make the Mott insulator, and the second one with $\rho_{B}$ is the extra charge by doping, so that the complex scalar describing the cooper pair condensation couples only with the second charge. We point out that the key role in creating the dome is played by the three point interaction $-c \chi^{2} F_{\mu\nu}G^{\mu\nu}$. The $Tc$ increases with their coupling. We also consider the effect of the quantum critical point hidden under the dome  using  the geometry of hyperscaling violation.  Our results  show that  the dome size and optimal temperature  increase with $z$ whatever is $\theta$,  while we get bigger $\theta$ for larger (smaller) dome depending on $z>2$ ($z<2$). We also point out  that the condensate increases for bigger value of $\theta$ but for smaller value of $z$.

keywords: superconducting dome,  quantum critical points, hyperscaling violating geometry
\end{abstract}

%\arxivnumber{2009.00381}
\end{center}

\maketitle

\section{Introduction}
Holographic duality, also known as the gauge/gravity duality  \cite{Maldacena:1997re,Gubser:1998bc,Witten:1998qj}, has brought many  insights in understanding strongly interacting electron systems, especially for condensed matter physics.  For example, the way to calculate transports and spectral feature of strongly correlated system has been suggested and new mechanism of  superconductivity has been suggested. See ref. \cite{Hartnoll:2016apf,zaanen2015holographic} and references therein.

\begin{figure}[!t]
\begin{centering}
\subfigure[two charges]
{\includegraphics[scale=0.2]{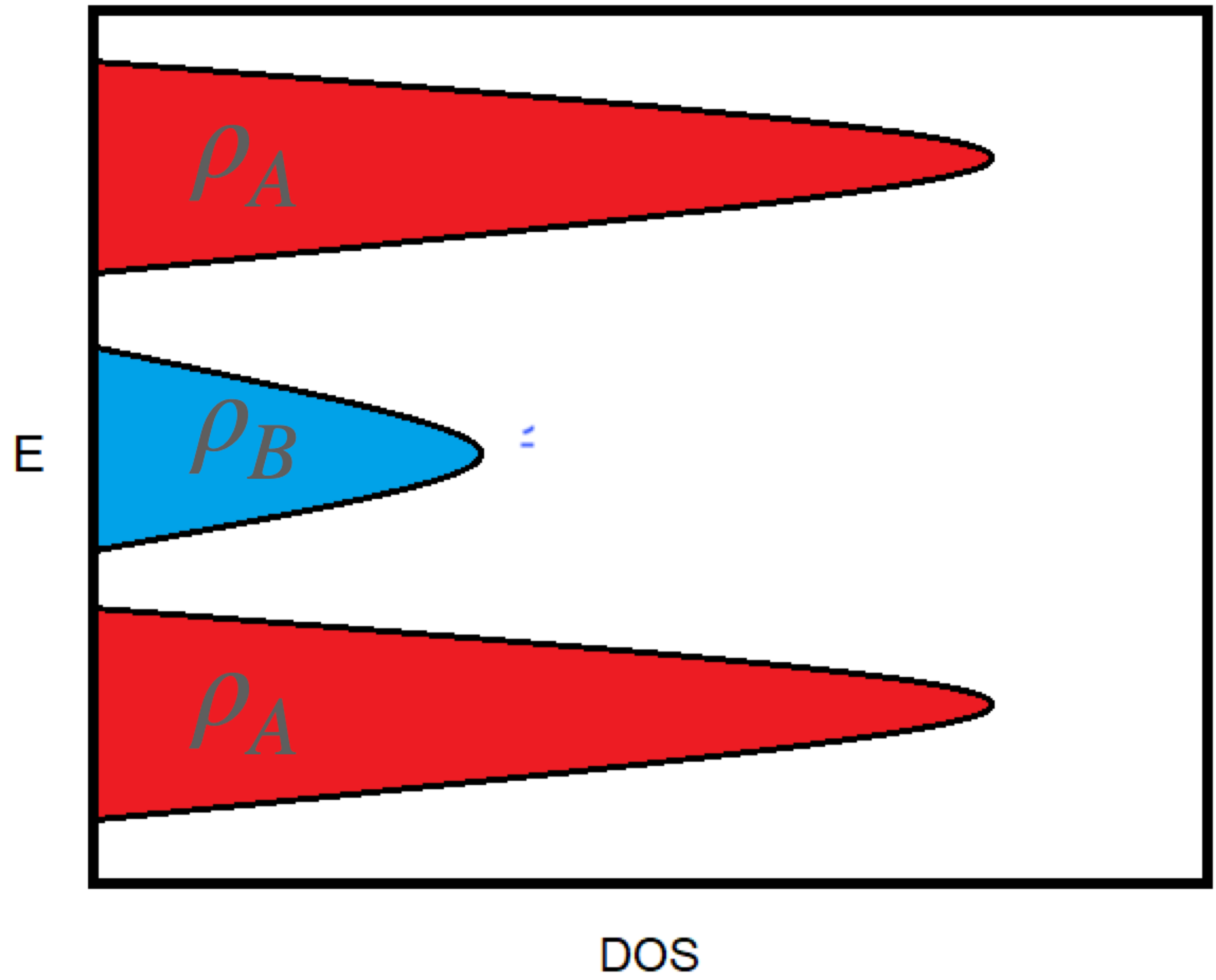}}
\subfigure[Phase diagram]
{\includegraphics[scale=0.46]{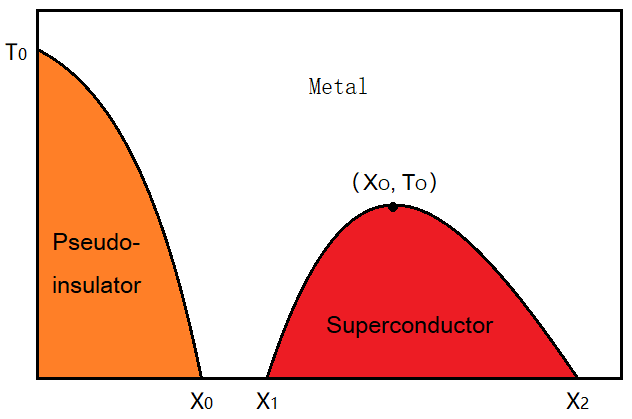}}
 \subfigure[ pseudo-insulator]
{\includegraphics[scale=0.44]{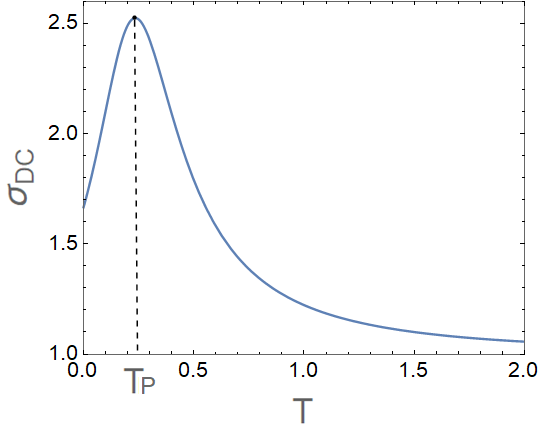}}
\par\end{centering}
\caption{\label{fig:Figure1} (a) The density of the states for doped Mott insulator. The degrees of freedom in the red region are provided by non-moving charge with density $\rho_{A}$, while those of the blue region is provided by the doped charges  $\rho_{N}$.
 (b) Phase diagram in the temperature-doping plane. The line connecting $(x_0,0), (0,T_0)$ denotes the phase boundary of the pseudo-insulating phase. $(x_1,x_2)$  is the interval of the superconducting dome. $(x_O,T_O)$ denotes the optimal $T_c$.
 (c)  The metal pseudo-insulator transition happens when the peak-position $T_p$ is at 0.  Here we used $\alpha=1$ and $x=3$. }
\end{figure}

 One important problem is to understand the phase diagram of the high $T_c$ superconductivity: various high-$T_c$ compounds all fit into a universal phase diagram where normal, superconducting, anti-ferromagnetic and pseudogap phases  compete \cite{LNW} and coexist. Recently, a holographic model was suggested  which qualitatively realizes the phase diagram  in the temperature-doping plane \cite{Kiritsis:2015hoa}.
 A simplified model was proposed in \cite{Baggioli:2015dwa} where the non-abelian gauge field  is replaced by a higher power of  scalar field that breaks the  translation symmetry. Here metal insulator transition is still realized although anti-ferromagnetic origin of the insulating phase is undermined.  In this model the metallic phase was  characterized by the DC conductivity decreasing with temperature, and the pseudo-insulator characterized by the DC conductivity increasing with temperature  (see the phase diagram in Figure \ref{fig:Figure1}(b).

In the previous papers \cite{Kiritsis:2015hoa,Baggioli:2015dwa}, the mobile carriers density is $\rho_A$, and $\rho_B=x\rho_A$ represents the impurities density. In this paper, we reinterpret the model such that it is more consistent with the doped Mott insulator. At zero doping, the lattice is half filled so that it has a Mott gap. The "Free" electrons can not move around due to the strong Coulomb repulsion. We will make sure the  $\rho_A$  electric charges do not move by assuming that,  in the conductivity calculation, the $\rho_A$ charge does not respond to the external field while $\rho_B$ charge responds. Then we dop the system with doping ratio $x$. For each dopant atom we can assume there are $1+1$ valence electrons: one contribute to the non-moving half filled state and the other contribute to itinerant electron whose number is say, $N_1=xN$. These are the movable electrons. Therefore the density of the non-movable charge density is always ( before and after doping,) $N/V=\rho_A$, and the movable charge density is $\rho_B=N_1/V=xN/V=x\rho_A$. We assume that as illustrated in Figure \ref{fig:Figure1} (a), $\rho_A$ only contributes to the electrons in Mott insulator, so that it does not contribute to conductor or superconductor. On the other hand, the doped charges of density $\rho_B$ contribute to the density of state near the Fermi surface, so that when
$x=\rho_{B}/\rho_{A}$ is large enough, charge from the impurities begins to make the superconductor. With this setup,
 we   studied    the roles  of the couplings on the phase diagram and  evaluated parameteric  dependence of the critical temperature   at the optimal doping.
 It turns out that what makes the superconducting  dome is the three  points coupling between the  density of cooper pair and two kind of  charges, namely
 \be
 {\cal L}_{int}= -c \chi^{2} F_{\mu\nu}G^{\mu\nu},
 \ee
where $\chi$ is the amplitude of the complex scalar describing the cooper pair condensation, and $F$ and $G$ are the field strengths of the two gauge fields created by the two kinds of charges.
Increasing the coupling $c$ makes the dome higher and  bigger,   therefore  increasing this coupling  is the key to increases the $T_{c}$.
Also we find that  by increasing the coupling $b$  of  the interaction term  $ {\cal L}_{int}= b \chi^{2} G_{\mu\nu}G^{\mu\nu},$ one can make the width of the dome smaller to resemble the cuperate  case.

 Since the superconducting dome is believed to cover a quantum critical points (QCP),  it is very interesting to explore the effects of dynamical exponents of the QCP.  The  hyperscaling violating geometry is precisely the geometry that realizes the   symmetry of a general class of  QCP.
We generalize the model of \cite{Baggioli:2015dwa} to a holographic model with hyperscaling violating geometry using the solution \cite{Ge:2016sel}, where  a black hole solution is derived with a dynamic exponent $z$ and a hyperscaling violating exponent $\theta$.
 We investigate the exponent dependence of the dome size as well as that of the superconducting condensate.
We found  that  the dome size and optimal temperature increase with $z$ whatever is $\theta$,  while we get larger dome for bigger $\theta$    if  $z>2$ and vice versa.
We also point out  that the condensate increases for bigger value of $\theta$ but for smaller value of $z$.

 The paper is organized as follows. In section 2, we first review the doped holographic superconductors with broken translational symmetry. Then, we investigate impact of the coupling  on the phase boundary,   on the endpoints of the superconducting dome and on the optimal $T_c$. We single out the key parameter to create the dome: the three points coupling between the density of cooper pair and  two different charge carriers. In section 3, in order to explore the effects of QCP, we extend M. Baggioli and M. Goykhman's model to the doped holographic superconductor with hyperscaling violation. In section 4, we focus on the superconducting condensate for $\theta=0$ and $\theta=1$. The section 5 is the summary and discussion. In the Appendix A, we show that two methods to obtain the superconducting dome are equivalent. In the Appendix B, we figure out the role of parameters $a,b,\alpha$ in the superconducting dome.

\section{Coupling dependence of the critical temperature}
We first briefly review the doped holographic superconductors with broken translational symmetry. Baggioli and Goykhman introduced a doped holographic model with a momentum dissipation \cite{Baggioli:2015dwa}, and find the superconducting dome. There are two U(1) gauge fields $A_\mu$ and $B_\mu$, two neutral scalar $\phi^I=\alpha x^I,(I=x,y)$, the complex scalar field $\psi=\chi e^{i\theta}$. $A_\mu=(A_t(u),0,0,0)$ is the bulk dual of the density of the charge carrier. $B_\mu=(B_t(u),0,0,0)$ is the bulk dual of the density of impurity. The neutral and massless scalars are responsible for the breaking of translational symmetry. The complex scalar field represents the order parameter for superconducting phase transition. $x=\rho_B/\rho_A$ is called as dope parameter. $u$ denotes holographic coordinate which is dual to the renormalization scale. The action is written as
 \begin{align}
S=\frac{1}{16\pi}\int d^4x \sqrt{-g}\left(R+\frac{6}{L^2}+{\cal L}_c+{\cal L}_s\right)
\end{align}
\begin{align}
{\cal L}_c&=-\frac{Z_A(\chi)}{4}A_{\mu\nu}A^{\mu\nu}
-\frac{Z_B(\chi)}{4}B_{\mu\nu}B^{\mu\nu}
-\frac{Z_{AB}(\chi)}{2}A_{\mu\nu}B^{\mu\nu}\\
&-\frac{1}{2}(\partial_\mu \chi)^2-H(\chi)(\partial_\mu \theta-q_AA_\mu-q_B B_\mu)^2
-V_{int}(\chi)\\
{\cal L}_n&=-2m^2V(X)\,.
\end{align}
We consider two interesting choices for $(q_A,q_B)$: $(1,0)$ or $(0,1)$. However, the density $\rho_A$ only contributes to the Mott insulator, so $(0,1)$ is the better choice for the couplings between the gauge fields and the complex scalar. Here
\begin{equation}
H(\chi)=\frac{n\,\chi^2}{2}\,,Z_A(\chi)=1-\frac{a\,\chi^2}{2}\,,Z_B(\chi)=1-\frac{b\,\chi^2}{2}\,,Z_{AB}(\chi)=\frac{c\,\chi^2}{2}\label{expansions}\,.
\end{equation}
\begin{align}
V_{int}(\chi)&=\frac{M^2\chi^2}{2}\,.
\end{align}
In this paper, we consider the non-linear Lagrangian
\begin{equation}
V(X)=\frac{X}{2m^2}+(\frac{X}{2m^2})^5\,,\,X=\frac{1}{2}g^{\mu\nu}\partial_\mu\phi^I\partial_\nu\phi^I\,.
\end{equation}
The corresponding equations of motion are as follows:
\begin{align}
&R_{\mu\nu}-\frac{Z_A}{2}A_{\nu\sigma}A_\mu^{\sigma}-\frac{Z_B}{2}B_{\nu\sigma}B_\mu^{\sigma}-\frac{Z_{AB}}{2}(A_{\nu\sigma}B_\mu^{\sigma}+A_{\mu\sigma}B_\nu^{\sigma})\nonumber\\
&-\frac{1}{2}\partial_\mu\chi\partial_\nu\chi-H(\partial_\mu\theta-q_AA_\mu-q_BB_\mu)(\partial_\nu\theta-q_AA_\nu-q_BB_\nu)-\frac{1}{2}g_{\mu\nu}L=0\,,\nonumber\\
&\nabla_\nu(Z_AA^{\nu\mu}+Z_{AB}B^{\nu\mu})+2q_AH(\nabla^{\mu}\theta-q_AA^\mu-q_BB^\mu)=0\,,\nonumber\\
&\nabla_\nu(Z_BB^{\nu\mu}+Z_{AB}A^{\nu\mu})+2q_BH(\nabla^{\mu}\theta-q_AA^\mu-q_BB^\mu)=0\,,\nonumber\\
&\nabla_\mu\nabla^\mu\chi-\frac{1}{4}\partial_\chi Z_A A^2-\frac{1}{4}\partial_\chi Z_B B^2-\frac{1}{2}\partial_\chi Z_{AB}A\cdot B-\partial_\chi H(\partial_\mu\theta-q_A A_\mu-q_B B_\mu)^2-\partial_\chi V_{int}=0\,.\nonumber
\end{align}
We obtain the solutions by solving the background equations of motion,
\begin{align}
&ds^2=\frac{L^2}{u^2}\left(-f(u)e^{-\tau(u)}dt^2+dx^2+dy^2+\frac{du^2}{f(u)}\right)\,,\\
&f(u)=\frac{(u-u_h) \left(2 u^2 \left(\alpha ^2 L^2 u_h^2-2\right)+u^3 u_h^3 \left(\rho_A^2+\rho_B^2\right)-4 u u_h-4 u_h^2\right)}{4 u_h^3}\,,\\
&A_t(u)=\rho_A(u_h-u)\,,\,B_t(u)=\rho_B(u_h-u)\,.
\end{align}
We emphasize that we take $(q_{A},q_{B})=(0,1)$ in this paper. But for the comparison, we also calculate for the   $ (1,0)$ case.
There are three phases on the temperature and doping plane: superconducting, metallic and pseudo-insulating (see the phase diagram in \cite{LNW}). As a generalization and comparison
to the  results in \cite{Baggioli:2015dwa}, it is interesting to discuss how much impact of the parameters $a,b,c,\alpha$ on $x_0,T_0,x_1,x_2,x_O,T_O$.

The DC conductivity can be calculated analytically following \cite{Baggioli:2014roa,Seo:2016vks} and the result is given by:
\begin{equation}
\sigma_{DC}=1+\frac{\rho_B^2 u_h^2}{2\alpha^2(1+5\frac{u^8_h\alpha^8}{(2m^2)^4 })}\,.
\end{equation}
Here $u_h$ is the location of the horizon.  Notice that we regards the $\rho_{A}$ is not moving degree of freedom and does not contribute the the conductivity.
There is other  way to introduce  insulator behavior by  the coupling   X  with the field strength \cite{Baggioli:2016oqk,An:2020tkn}.  Here, metal insulator transition is not our main concern. Following \cite{Blake:2013bqa,Donos:2014cya,Grozdanov:2015qia}, we obtain the DC conductivity. The pseudo-insulating phase by the change from metallic ($d\sigma_{DC}/dT<0$) to pseudo-insulating ($d\sigma_{DC}/dT>0$) is shown in Figure \ref{fig:Figure1} (c). We vary the ratio $x$ and evaluate the movement of the peak-position. Our results imply that $T_p$ decreases as $x$ increases.

Now, we discuss the instability of a normal phase of the bulk system to determine whether a boundary system has a superconducting phase. The effective mass is read off where BF bound is violated. We first derive instability analytically in the normal state at zero temperature. Two endpoints $x_1$ and $x_2$ can be determined by the violated $AdS_2$ BF bound, namely $m^2_{eff}L^2_2<-1/4$.

We furthermore study the superconducting dome in a finite-temperature normal phase background. We begin with the lineared equation of motion for $\chi$. The scalar $\chi$ represents the order parameter for superconducting phase transition (see more details in \cite{Kiritsis:2015hoa,Denef:2009tp}). Then we look for the finite temperature black brane solution to determine the superconducting dome, which satisfies the two boundary conditions. It is regular at the horizon, so it demands the expansion as follow
\begin{align}
&f(u)=f'(u_h)(u-u_h)+\mathcal{O}(u-u_h)^2\,,\nonumber\\
&\chi(u)=\chi(u_h)+\chi'(u_h)(u-u_h)\,,\,\tau(u)=\tau(u_h)+\tau'(u_h)(u-u_h)\,,\nonumber\\
&A_t=\rho_A(u_h-u)\,,\,B_t=\rho_B(u_h-u)\,.\nonumber
\end{align}
We solve the lineared equation for $\delta\chi$, which is considered as a probe on the AdS-RN background
\begin{align}
 &\delta\chi ''(u)+\left(\frac{f'(u)}{f(u)}-\frac{2}{u}\right) \delta\chi '(u)\nonumber\\
 &+\frac{f(u) \left(u^4 (a+x (b x+2 c))-2 M^2\right)+2 n u^2 (u-u_h)^2 (q_A+q_B x)^2}{2 u^2 f(u)^2}\delta\chi(u)=0\,.
\end{align}
There are five independent parameters $u_h,\chi(u_h),\tau(u_h),\rho_A,\rho_B$. This scaling symmetry can be used to set $\tau(u_h)=0$, and we also set $\rho_A=1\,,\,\rho_B=x$. Actually, we are left with two independent parameters $x$ and $u_h$, namely doping and temperature. By instituting into equations of motion for $\delta\chi$, we solve background equations of motion numerically out to large value of $r$. Based on this initial condition. Then, we obtain the source free solution from horizon to the boundary. Afterwards, we keep the doping parameter $x$ fixed, and find the maximal $T_c$ (see Appendix A for details).

\begin{figure}[!t]
\begin{centering}
%\subfigure[definition of pseudo-insulator]
{\includegraphics[scale=0.4]{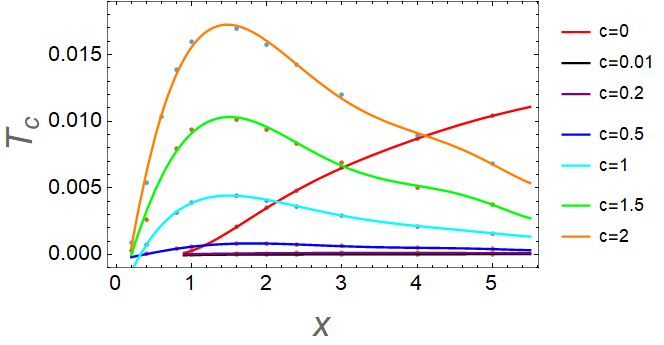}}
\par\end{centering}
\caption{\label{fig:Figure 2} Role of $c$. We  plot the superconducting phase with different $c=0,0.01,0.2,0.5,1,1.5,2$ $a=b=0$. No dome without $c$.  Larger dome with increasing $c$.  Notice that $c\to 0$ is a singular limit. }
\end{figure}

We focus on the three points coupling between the density of cooper pair and two kind of charges: $-c \chi^{2} F_{\mu\nu}G^{\mu\nu}$, because the coupling $c$ is the key parameter for the superconducting dome. We first discuss the role of $c$ in the simplified model
\begin{equation}
Z_A(\chi)=1\,,Z_B(\chi)=1\,,Z_{AB}(\chi)=\frac{c\,\chi^2}{2}\,.
\end{equation}
Our results are presented in Figure \ref{fig:Figure 2}. If c is negative, there is no instability solution. In this paper, we do not need to consider the negative c. The superconducting dome occurs if $c\neq 0$ and the dome expands as $c$ increases. Our result shows that the superconducting expands as $c$ increase. If the cross term vanishes $(c=0,Z_{AB}=0)$, there is no superconducting instability. A dome exists even at $a=b=0$ if $c\neq0$. Our result shows that $c$ alone is almost enough to create the dome. Therefore, the coupling $c$ is crucial to high $T_c$.

\begin{figure}[!t]
%\begin{multicols}{2}
 \begin{center}
 $\,$\\
 $\,$\\
 $\,$\\
 \subfigure[ Role of $c$ ]
 {\includegraphics[scale=0.53]{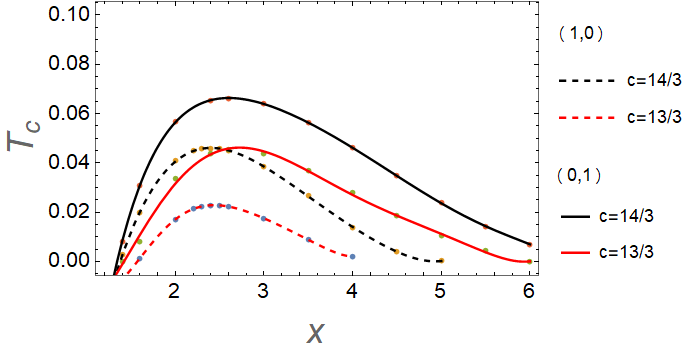}}
  \subfigure[ left endpoint $x_1(c)$ ]
 {\includegraphics[scale=0.25]{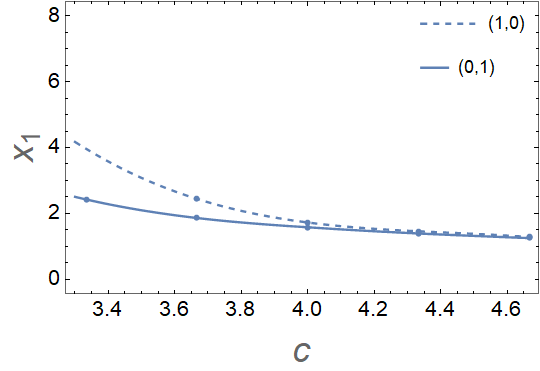}}
 \subfigure[ right endpoint $x_2(c)$ ]
 {\includegraphics[scale=0.25]{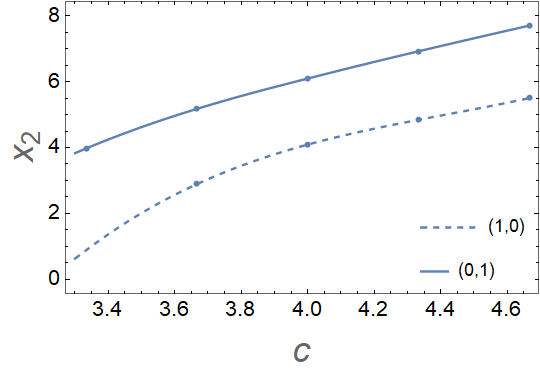}}
  \end{center}
%  \end{multicols}
  \caption{\label{fig:Figure 3}  Role of $c$ : (a)The dome-shaped superconducting phase with the the coefficient $c$ appearing in the expansion (\ref{expansions}).  (b,c) The left endpoint $x_1$ and the right endpoint $x_2$ of the superconducting region. We fix $a=10, b=4/3, \alpha=0$ and choose different $c$. The dashed line represents $(1, 0)$, and the solid line represents $(0, 1)$.}
\end{figure}

Then, we discuss the case of $a\neq 0\,,\,b\neq 0$. We find that the behavior of endpoints is the same for $(1, 0)$ or $(0, 1)$. Especially,  $x_1$ moves left and $x_2$ moves right when $c$ increase to make the dome bigger (see Figure \ref{fig:Figure 3}).
The optimal temperature $T_O$ with $(q_A,q_B)=(1,0)$ is always smaller than the case of $(0,1)$. This result is not surprising  because %$T_c\sim e^{-1/(Ng)}$.
the coupling $Z_B=1-\frac{b\chi^2}{2}$ is stronger than the coupling $Z_A=1-\frac{a\chi^2}{2}$ in our choice. By the same reason, the superconducting dome of $(0,1)$ theory is bigger than the one with $(1,0)$ one. The roles of other paramters are given in the Appendix B. Perhaps the most interesting result is the fact that the coupling described by $c$ alone can generate the superconducting dome and increasing its size, and the roles of other coupling $Z_{A}$ $Z_{B}$ are secondary ones for producing the dome.

\section{The doped superconductor with hyperscaling violation}
In the previous section, we studied the doped holographic superconductor with broken translational symmetry. In order to explore the effects of the QCP, it would be natural to extend the present analysis to the case of hyperscaling violation. We begin with the black hole solutions in Einstein-Maxwell-axion-dilaton theory with a dynamic exponent $z$ and a hyperscaling violation exponent $\theta$ \cite{Ge:2016sel}, where a $U(1)$ gauge field is considered as an auxiliary gauge field, leading to a Lifshitz-like vacuum. The neutral and massless scalars $\phi^I\ :\ \phi^x=\alpha x\ ,\ \phi^y=\alpha y$ generate momentum relaxation. We generalize this model by introducing a perturbative charge scalar $\psi=\chi e^{i\phi}$ in four-dimensional bulk spacetime. Now there are three gauge fields. $A_1$ is the auxiliary gauge field to support the Lifshitz gravity as usual, while $A_2$ and $A_3$ are the physical gauge fields which provide the finite chemical potentials.
$A_2$ is the bulk dual of the density of the charge carrier $\rho_A$ which describe the  Mott insulator while $A_3$  is the bulk dual of the density of doped charge $\rho_B$ which can move and provides the density of state at the Fermi surface. We describe the non-moving nature as the absence of the coupling of such charge with the electric field. Therefore our model is $q_{B}=1, q_{A}=0$. To compare with previous treatment, we also considered results coming from the $q_{A}=1, q_{B}=0$.

\begin{figure}[!t]
\begin{centering}
\includegraphics[scale=1.2]{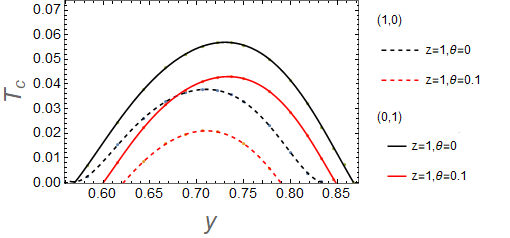}
\par\end{centering}
\caption{\label{fig:Figure 4}Phase diagram on the temperature-doping plane with different $\theta$ and fixed $z = 1$. In this figure, solid lines are for $(q_A,q_B)=(1,0)$, and dashed lines are for $(q_A,q_B)=(0,1)$. Black and red $\theta = 0, 0.1$, respectively. The critical temperature decreases as $\theta$ increases.}
\end{figure}

\begin{figure}[!t]
\begin{centering}
\includegraphics[scale=1.2]{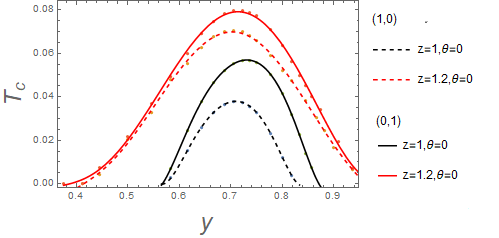}
\par\end{centering}
\caption{\label{fig:Figure 5}Phase diagram on the temperature-doping plane with different $z$ and fixed $\theta = 0$. In this figure, solid lines are for $(q_A,q_B)=(1,0)$, and dashed lines are for $(q_A,q_B)=(0,1)$. Black and red represent $z = 1, 1.2$, respectively. The critical temperature increases as $\theta$  increases.}
\end{figure}

\begin{figure}[!t]
\begin{centering}
\includegraphics[scale=1.3]{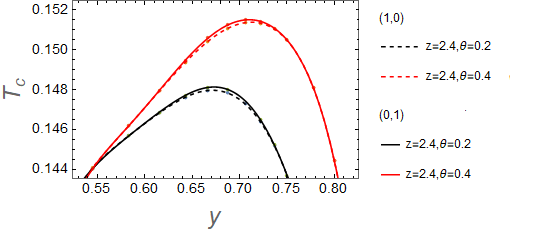}
\par\end{centering}
\caption{\label{fig:Figure 6}Phase diagram on the temperature-doping plane with different $\theta$ and fixed $z = 2.4$. In this figure, solid lines are for $(q_A,q_B)=(1,0)$, and dashed lines are for $(q_A,q_B)=(0,1)$. Black and red represent $\theta = 0.2, 0.4$, respectively. The critical temperature increases as $z$ increases.}
\end{figure}

\begin{figure}[!t]
\begin{centering}
\includegraphics[scale=0.38]{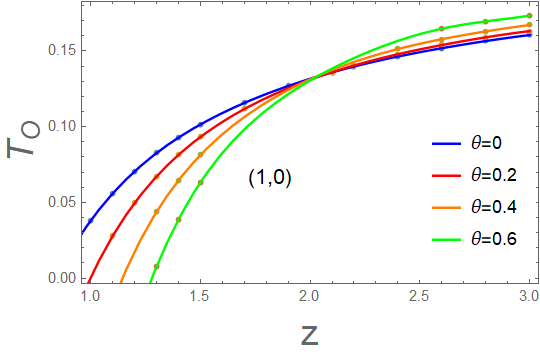}\includegraphics[scale=0.37]{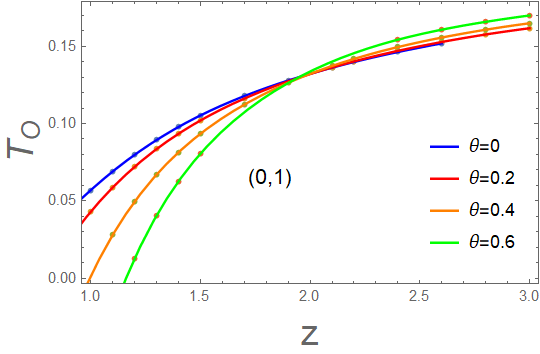}
\par\end{centering}
\caption{\label{fig:Figure 7}Relation between the optimal temperature $T_O$ and $z,\theta$. The maximum critical temperature defines the optima doping. Left: $(q_A,q_B)=(1,0)$. Right: $(q_A,q_B)=(0,1)$.}
\end{figure}

The new doping parameter is defined as
\begin{equation}
  y=\frac{x}{1+x}\,,\,x=\rho_B/\rho_A\,.
\end{equation}
Here $0<y<1$. Since $x$ is large in the superconducting dome, we consider $y$ instead of $x$ in the following calculation. The action is written as
\begin{eqnarray}
  S&=&\frac{1}{16\pi}\int d^{4}x\sqrt{-g}\big[R+\frac{6}{L^2}-\frac{1}{4}(e^{\lambda_1\phi})F_1^2-\frac{1}{4}(e^{\lambda_2\phi}+\frac{a\chi^2}{2})F_2^2\nonumber\\
  &-&\frac{1}{4}(e^{\lambda_3\phi}+\frac{b\chi^2}{2})F_3^2-\frac{1}{2}\frac{c\chi^2}{2}F_2\cdot F_3-\frac{1}{2}(\partial\chi_{i})^{2}+H(\chi)(\partial_\mu\theta-q_A A_2-q_B A_3)^2\nonumber\\
  &-&V_{int}(\chi)-\frac{1}{2}(\partial\phi)^{2}-\frac{1}{2}e^{\eta\phi}(\partial\phi^I)^2+V e^{\gamma\phi}\big]
\end{eqnarray}
Here we define the following coupling and potential
\begin{equation}
 H(\chi)=\frac{n\chi^2}{2}\ ,\ V_{int}(\chi)=\frac{M\chi^2}{2}\,.\nonumber\\
\end{equation}
Following \cite{Ge:2016sel}, we have
\begin{align}
 &q_1=\sqrt{2}\sqrt{-2+z+z^2+\theta-z\theta}\nonumber\\
 &Z_1=e^{\lambda_1\phi}=u^{-(\theta-4)}\ ,\ \ Z_2=e^{\lambda_2\phi}=u^{(\theta-2z+2)}\ ,\ \ Z_3=e^{\lambda_3\phi}=u^{(\theta-2z+2)}\nonumber\\
 &Y=e^{\eta\phi}=\frac{1}{Z_2}\ ,\ \ \phi=\sqrt{(2-\theta)(2z-2-\theta)}\ln\frac{1}{u}\ ,\ \ V=(z-\theta+1)(z+2-\theta)u^{-\theta}\nonumber
\end{align}
The matric ansatz is
\begin{align}
  &ds^2=\frac{1}{u^{2-\theta}}\bigg(-\frac{f(u)}{u^{2(z-1)}}e^{-\tau(u)}dt^2+dx^2+dy^2+\frac{du^2}{f(u)}\bigg)\,,\nonumber\\
  &A_1=a_1(u)dt\ ,\ A_2=a_2(u)dt\ ,\ A_3=a_3(u)dt\,,\nonumber\\
  &\chi=\chi(u)\ ,\ \theta\equiv 0\,.
\end{align}
where $u$ is the radial bulk coordinate. We obtain the solution as follows,
\begin{align}
  &f(u)=1-\frac{m}{u^{\theta-z-2}}-\frac{\alpha^2}{(\theta-2)(z-2)u^{\theta-2z}}+\frac{(\rho_A^2+\rho_B^2)(\theta-2)}{2(\theta-2)u^{2\theta-2z-2}}\\
  &m=-\frac{1}{u_h^{2+z-\theta}}\big(-1+\frac{\alpha^2}{(-2+z)(-2+\theta)u_h^{-2z+\theta}-\frac{(\rho_A^2+\rho_B^2)(-z+\theta)}{2(-2+\theta)u_h^{-2-2z+2\theta}}}\big)\\
  &a_1(u)=\frac{q_1}{2+z-\theta}(\frac{1}{u^{2+z-\theta}}-\frac{1}{u_h^{2+z-\theta}})\nonumber\\
  &a_2(u)=\mu_A-{\rho_A}{u^{z-\theta}}\ ,\ a_3(u)=\mu_B- {\rho_B}{u^{z-\theta}}\,,\nonumber
\end{align}
When $z=1,\,\theta=0$, $f(u)$ return to the form in \cite{Baggioli:2015dwa}. Following \cite{Kiritsis:2015hoa}, we fix $a=10,\,b=4/3,\,c=14/3,\,n=1,\,\alpha=1,\,M^2=-5/4$. The superconducting dome survives the translational symmetry breaking \cite{Baggioli:2015dwa}. Our calculation shows the result still holds even for hyperscaling violating background. When $z=1,\,\theta=0$, our result could return to the one in \cite{Baggioli:2015dwa}. The phase diagrams with different $z$ and $\theta$ is presented in Figure \ref{fig:Figure 4}-Figure \ref{fig:Figure 6}.
The numerical optimal temperatures $T_O$ with different $z$ and $\theta$ are presented in Table 1.
\begin{sidewaystable}
\begin{tabular}{|c||c|c|c|c|c|c|c|c|c|c|c|c|c|c|c|c}
\hline $z\ \backslash\ \theta$ &1&11/10&12/10&13/10&14/10&15/10&16/10&17/10&19/10&21/10&22/10&24/10&26/10&28/10&3\\
\hline\hline
-1/10&0.051&0.066&0.078&0.088&0.097&$=$&$=$&-&-&-&-&-&-&-&-\\
\hline 0&0.038&0.056&0.071&0.083&0.093&0.102&-&0.116&0.127&0.136&0.140&0.146&0.152&0.157&0.161\\
\hline 1/10&0.021&0.044&0.061&0.076&0.093&0.098&-&0.114&0.126&0.136&0.140&0.147&0.153&0.158&0.162\\
\hline 2/10&-&0.028&0.050&0.067&0.081&0.093&-&0.112&0.126&0.136&0.141&0.148&0.154&0.159&0.163\\
\hline 3/10&-&0.008&0.035&0.057&0.074&0.088&-&0.109&0.126&0.137&0.141&0.149&0.156&0.161&0.165\\
\hline 4/10&-&-&-&0.044&0.065&0.081&0.095&0.093&$\circ$&$\circ$&$\circ$&0.151&0.158&0.163&0.167\\
\hline 5/10&-&-&-&0.028&0.053&0.073&0.090&$\circ$&$\circ$&$\circ$&$\circ$&0.154&0.161&0.166&0.170\\
\hline 6/10&-&-&-&0.008&0.039&0.063&0.083&$\circ$&$\circ$&$\circ$&$\circ$&$\circ$&0.165&0.170&0.173\\
\hline 7/10&-&-&-&-&0.022&0.051&$\circ$&$\circ$&$\circ$&$\circ$&$\circ$&$\circ$&$\circ$&0.174&0.178\\
\hline 8/10&-&-&-&-&-&0.037&$\circ$&$\circ$&$\circ$&$\circ$&$\circ$&$\circ$&$\circ$&$\circ$&0.184\\
\hline 9/10&-&-&-&-&-&0.023&0.053&$\circ$&$\circ$&$\circ$&$\circ$&$\circ$&$\circ$&$\circ$&$\circ$\\
\hline 1&$\bullet$&$\bullet$&0.028&-&-&$\bullet$&0.043&$\circ$&$\circ$&$\circ$&$\circ$&$\circ$&$\circ$&$\circ$&$\circ$\\
\hline
\end{tabular}
\caption{Optimal temperature $T_O$ and SC dome with different z and $\theta$: ``-'' denotes negative $T_O$. ``$\bullet$'' denotes the monotonously decreasing $T_c$. ``$\circ$'' denotes the monotonously increasing $T_c$. ``$=$'' denotes that there is no instability solution. Here we fix $(1\,,\,0)$.}
\end{sidewaystable}
The relation between the optimal temperature $T_O$ and dynamical $z$ with different $\theta$ is shown in Figure \ref{fig:Figure 7}. It shows that $T_O$ increase as z increase. The result is in agreement with the numerical calculation without the mass of the probed scalar field in \cite{Zhang:2015dyz}. $T_O$ decrease as $\theta$ increases when $z<2$. However, we have the opposite behavior when $z>2$. Figure \ref{fig:Figure 4} and Figure \ref{fig:Figure 6} also indicate that for $z<z_c$ the optimal $T_c$ decreases as $\theta$ increases, while for $z>z_c$, $T_O$ increases as $\theta$ increases. The result is not in agreement with the previous work \cite{Zhang:2015dyz,Pan:2015lit,Fan:2013tga}. The difference is expected  because our model is quite different from their model. There is only one gauge field in their model, so superconducting dome can not be generated.

\section{Condensate  for  hyperscaling violation}
\begin{figure}[!t]
\begin{centering}
\includegraphics[scale=0.7]{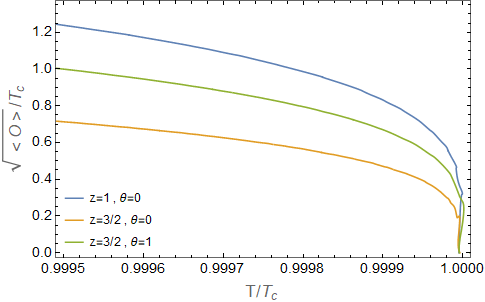}
\par\end{centering}
\caption{\label{fig:Figure 8}The cooper pair condensate $<\mathcal{O}>$ with fixed $\alpha=1\,,\,x=2$. The blue line is for $z=1,\theta=0$. The orange line is for $z=3/2,\theta=0$. The green line is for $z=3/2,\theta=1$. The condensate increases  for bigger value of $\theta$, but it is suppressed by $z$.}
\end{figure}

We start with AdS black hole with dynamic exponent and hyperscaling violation exponent. The Hawking temperature  is determined by
\begin{equation}
T=\frac{f'(r_0)}{4\pi}\,.
\end{equation}
Therefore we define temperature as $T=r_0$. Consequently, we rewrite the equations of motion for $A_t,B_t,\chi$ in terms of the scaled fields: $A_t\rightarrow\frac{A_t}{r_0}\,,\,B_t\rightarrow\frac{B_t}{r_0}\,,\,\chi\rightarrow\frac{\chi}{r_0}\,,\,\mu\rightarrow\frac{\mu}{r_0}$. According to $\mu\rightarrow\frac{\mu}{r_0}=\frac{\mu}{T}$, fixing $\mu$ by changing $T$ is equivalent to fixing $T$ by changing $\mu$. In order to simplify the following computations, we fix $r_0=1$.
At the horizon, we demand the following expansion
\begin{align}
  &A_t=\alpha_0(r-1)+\alpha_1(r-1)^2+\alpha_2(r-1)^3\,,\nonumber\\
  &B_t=\beta_0(r-1)+\beta_1(r-1)^2+\beta_2(r-1)^3\,,\nonumber\\
  &\chi=\gamma_0+\gamma_1(r-1)+\gamma_2(r-1)^2\,.\nonumber
\end{align}
Meanwhile we can impose the boundary condition $A_t=B_t=0$, $A'_t=\alpha_0, B'_t=\beta_0$, $\chi=\gamma_0$.
The horizon regularity keeps conditions for equations of motion
\begin{align}
&Z_A\,(2\,A_t''+\tau'\,A_t')+Z_{AB}\,(2\,B_t''+\tau'\,B_t')+2\,\chi' (\dot{{Z}}_A\,A_t'+\dot{ Z}_{AB}\,B_t')
-4\,q_A\, H\,\frac{q_A \,A_t+q_B\, B_t }{u^2\,f}=0\,,\label{At}\\
&Z_B\,(2\,B_t''+\tau'\,B_t')+Z_{AB}\,(2\,A_t''+\tau'\,A_t')+2\chi' (\dot{Z}_B\,B_t'+\dot {Z}_{AB}\,A_t')
-4\,q_B\, H\,\frac{q_A \,A_t+q_B\, B_t }{u^2\,f}=0\,,\label{Bt}\\
&\chi''{+}\left(\frac{f'}{f}{-}\frac{2}{u}{-}\frac{\tau'}{2}\right)\chi'{-}\frac{L^2}{u^2\,f}\dot V_{int}
{+}\frac{e^\tau \,u^2}{2\,f}\left(\dot{\bar{Z}}_A\,\bar{A}_t^{\prime 2}{+}\dot{\bar{Z}}_B\,\bar{B}_t^{\prime 2}
{+}2\,\dot{Z}_{AB}\,A_t'\,B_t'\right)
{+}\frac{e^\tau \,\dot{H}}{f^2}\left(q_A\,A_t{+}q_B \,B_t\right)^2{=}0\,.\label{chi}
\end{align}
After solving the conditions for $\alpha_0,\beta_0,\gamma_0,\alpha_1,\beta_1,\gamma_1,\alpha_2,\beta_2,\gamma_2$, we are left with three independent parameters $\alpha_0,\beta_0,\gamma_0$. By integrating out to infinity, these solutions of (\ref{At}-\ref{chi}) for given $\alpha_0,\beta_0,\gamma_0$ behave as
\begin{align}
  &\chi=\frac{\chi_1}{r}+\frac{\chi_2}{r^2}\,,\nonumber\\
  &\chi_1=r\,\chi(r),\chi_2=-r^{2}(r\,\chi(r))'\,,\nonumber\\
  &A_t=\mu_A-\frac{\rho}{r}\,,\,B_t=\mu_B-\frac{\rho}{r}\,.\nonumber
\end{align}
We set $\chi_1=0$ and $\mu_A=1$ to determine a curve. Along this curve, we can calculate $\chi_2$ and $\mu_B$ at infinity \cite{Baggioli:2015zoa}. Somewhere along this curve $\chi_2$ could be zero. To determine critical point, $\mu_c$ is defined by $\chi_1=\chi_2=0$. Based on the approach in \cite{Hartnoll:2008vx,Hartnoll:2008kx}, we explore the superconducting condensate with the dynamic exponent $z$ and the hyperscaling violation exponent $\theta$. Here we vary $1/\mu$ instead of varying temperature, and study the superconducting condensate with $\theta=0$ for bosonic case and $\theta=1$ for fermionic case. The cooper pair condensate $<\mathcal{O}>$ presented in Figure \ref{fig:Figure 8} suggests that the couplings $Z_A$ and $Z_B$ are positive for all radial position in the case of broken translational symmetry and hyperscaling violation. The condensate becomes easier for bigger value of $\theta$, but it is suppressed by $z$. The behavior is in agreement with the result in \cite{Zhang:2015dyz,Pan:2015lit}. However, the superconducting dome expands and the optimal temperature increases when $z$ increases in Section 3. The results are not contradictory, since $<O>$ is not only a function of $T_c$ \cite{Siopsis:2010uq} but also the function of $z$ and $\theta$ \cite{Zhang:2015dyz,Lu:2013tza,Luo:2016ydt}. As z increases, the Fermi surface of cuprates becomes a Fermi arc, which means that it cannot be in a closed shape \cite{Fang:2012pw,Alsup:2014uca}. So the number of superconducting electrons decrease. Consequently superconducting condensate decreases.

\section{Conclusion and discussion}
In this paper, we have studied the holographic theory which has the ability to realize the doped high-temperature superconductors. Based on the previous work on pseudo-insulator and superconducting phases on the doping-temperature plane, especially instability conditions at zero temperature and at finite temperature, we further evaluate the impact of coefficients $a,b,c,\alpha$ on the corner region of pseudo-insulator and the superconducting dome with broken translational symmetry. Our results show the three points coupling $c$ is the key parameter. Afterwards, we extend our analysis to the case with a dynamic exponent and a hyperscaling violation exponent. Besides, we also calculate the superconducting condensate to verify the region of the couplings $Z_A\,,\,Z_B$.

We find which parameter has the most influence on the boundary of the pseudo-insulating phase ($x_0, T_0$), two endpoints $(x_1,x_2)$ of the superconducting dome and the optimal temperature $(x_O,T_O)$. As noted in Figure \ref{fig:Figure 10} and Figure \ref{fig:Figure 11}, we notice that the left endpoint $x_1$ is more sensitive to $a$, but the right endpoint $x_2$ is more sensitive to $b$. Furthermore, our result indicates that the superconducting dome is expanding as the couplings $Z_A,Z_B,Z_{AB}$ increase. The result is an advance in the research on high-temperature superconductivity.

In our calculation, $q_B$ can be chosen to be zero, so $\rho_B$ can be considered as the density of spin doping. One of the main progress of this paper is to show that the superconducting dome-shaped region with $(0,1)$ is much bigger than the one with $(1,0)$, and the three point interaction $-c \chi^{2} F_{\mu\nu}G^{\mu\nu}$ alone is almost enough to create the dome. Our numerical calculation also shows the superconducting dome survives when $\theta\neq0$.  Let us combine our results with \cite{Kiritsis:2015hoa,Baggioli:2015dwa}. The superconducting phase with hyperscaling violation is in qualitative agreement with the case of $\theta=0$. The depth of the superconducting dome increase as the dynamical exponent $z$ increase. The depth first decrease then increase as the hyperscaling violation exponent $\theta$ increase. So the other important progress of this paper is that the superconducting dome-shaped region built in \cite{Kiritsis:2015hoa} can be expanded by hyperscaling violation exponent $\theta$ if $z$ is large enough.

However, there are still some issues concerning the instability of our holographic setup in our calculations. First, gauge field $B_\mu$ is used to mimic the charge change due to doping. The rising part of the superconducting dome is attributed to the increasing of charge. So what about the descend part of the superconducting dome? Unfortunately, it seems hard to find out the reason of the descend part, since the Lagrangian indicates there is a symmetry between $A_\mu$ and $B_\mu$. Second, our numerical results about the superconducting dome imply that $y=\frac{1+x}{x}$ can approach to 0.85, namely $\rho_B\gg\rho_A$. It suggests doping charge is dominant in this case. We leave these issues to a future study.

There are much more generalized form of   hyperscaling violating solutions for black holes with  multiple fields in \cite{Li:2016rcv}.
 It would be interesting to study the consequence of the such solution along the line we studied here.

\section*{Appendix A: Methods to obtain the superconducting dome}

In this appendix, we verify two methods to obtain the superconducting dome are equivalent. There are two methods to determine the superconducting dome. One is to keep the doping parameter x fixed. Then we obtain the maximal temperature indicated by $\bullet$. The other method is to keep the horizon $u_h$ fixed. Then we obtain the maximal x indicated by $\blacklozenge$ and minimal $x$ indicated by $\vartriangle$. There is one-to-one match between the critical temperature $T$ and doping $x$.
\begin{figure}[!t]
\begin{centering}
\includegraphics[scale=0.6]{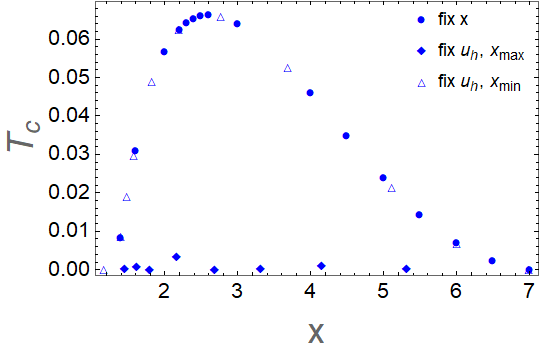}
\par\end{centering}
\caption{\label{fig:Figure 9} The dome-shaped region of the superconducting instability. We use two methods to obtain the critical temperature $T_c$ as a function of doping parameter $x$. Here we fix the parameters
as $M^2=-5/4, m=1, a = -10, b = -4/3, c = 14/3, n = 1,\rho_A=1, \alpha=0, \rho_B=x$ and $(0, 1)$.}
\end{figure}
 Figure \ref{fig:Figure 9} shows that these two methods are equivalent. The endpoints of the superconducting region are $x_1\approx 1.17$ and $x_2\approx 7$. The result indicates that temperature gradually approaches to zero when x increases. If $Z_A,Z_B,Z_{AB}$ are independent of $\chi$, the dome vanish and the critical temperature always increases. In order to simplify the calculations, we choose the former method in this paper.

\section*{Appendix B: Role of $a,b,\alpha$ in the superconducting dome}
\begin{figure}[!t]
%\begin{multicols}{2}
 \begin{centering}
 $\,$\\
 $\,$\\
 \subfigure[ Role of $a$ ]
 {\includegraphics[scale=0.54]{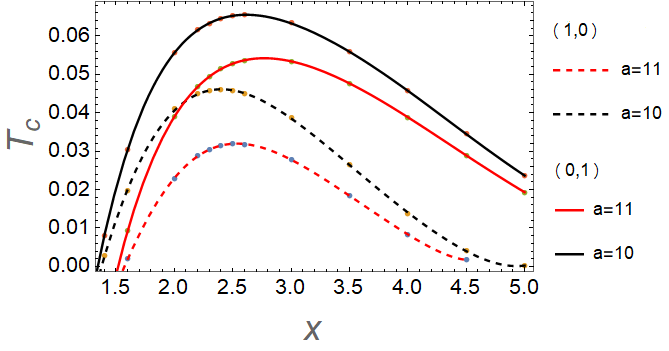}}
 \subfigure[ left endpoint $x_1(a)$ ]
 {\includegraphics[scale=0.25]{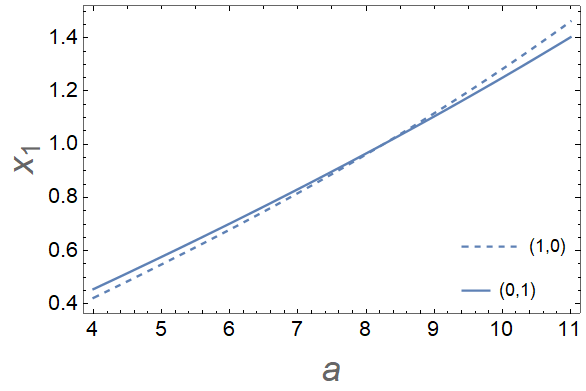}}
 \subfigure[ right endpoint $x_2(a)$ ]
 {\includegraphics[scale=0.24]{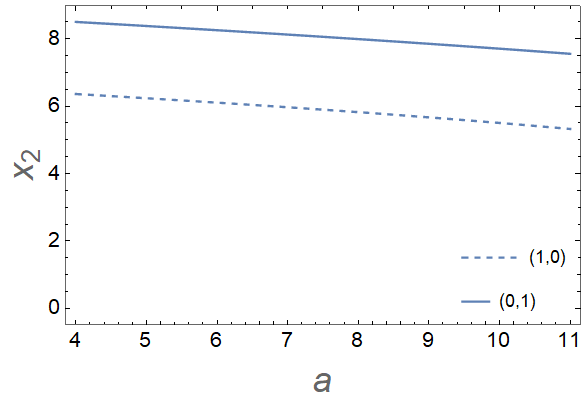}}
  \par\end{centering}
  %\end{multicols}
  \caption{\label{fig:Figure 10} Left side: The dome-shaped superconducting phase with the the coefficient $a$ appearing in the expansion (\ref{expansions}). Dome size increases for large $Z_A$. Right side: The left endpoint $x_1$ and the right endpoint $x_2$ of the superconducting region. We fix $b=4/3, c=14/3, \alpha=0$ and choose different $a$. The dashed line represents $(1,0)$, and the solid line represents $(0,1)$. We find $x_1(a)$ is increasing function.}
\end{figure}

\begin{figure}[!t]
%\begin{multicols}{2}
 \begin{center}
  $\,$\\
 $\,$\\
 \subfigure[ Role of $b$ ]
 {\includegraphics[scale=0.53]{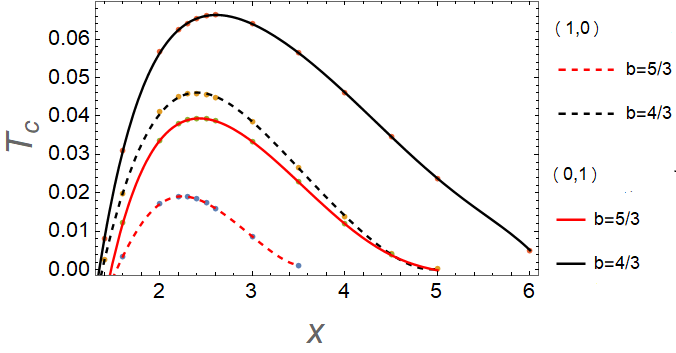}}
 \subfigure[ left endpoint $x_1(b)$ ]
 {\includegraphics[scale=0.26]{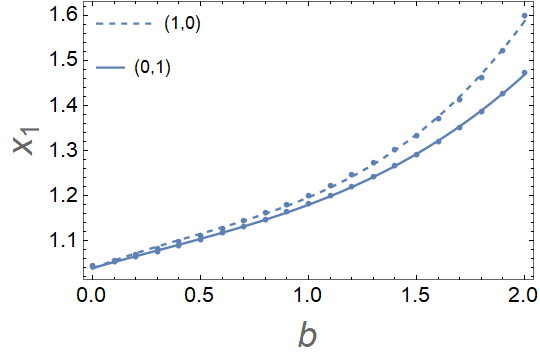}}
 \subfigure[ right endpoint $x_2(b)$ ]
 {\includegraphics[scale=0.26]{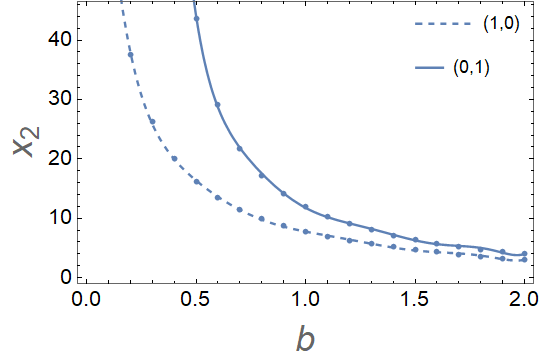}}
  \end{center}
  %\end{multicols}
  \caption{\label{fig:Figure 11} (a) The role of $b$.  The dome-shaped superconducting phase with the the coefficient $b$. Dome size increases for large $Z_B$.  (b,c)The left endpoint $x_1$ and the right endpoint $x_2$. We fix $a=10, c=14/3, \alpha=0$. The dashed line represents $(1, 0)$, and the solid line represents $(0, 1)$.}
\end{figure}

\begin{figure}[!t]
\begin{centering}
\subfigure[ $\alpha$ dependence of  $x_{0}$ ]
{\includegraphics[scale=0.3]{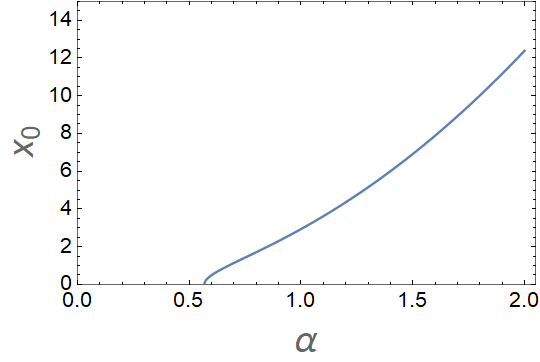}}\quad\quad
\subfigure[$\alpha$ dependence of  $T_{0}$]
{\includegraphics[scale=0.3]{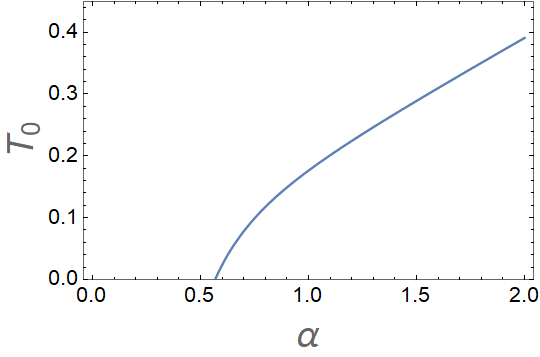}}
\par\end{centering}
\caption{\label{fig:Figure 12} The boundary of the pseudo-insulating phase $x_0$ and $T_0$ with the translational symmetry broken by the neutral scalars.}
\end{figure}

\begin{figure}[!t]
%\begin{multicols}{2}
 \begin{centering}
  $\,$\\
 $\,$\\
 \subfigure[ Role of $\alpha$ ]
 {\includegraphics[scale=0.54]{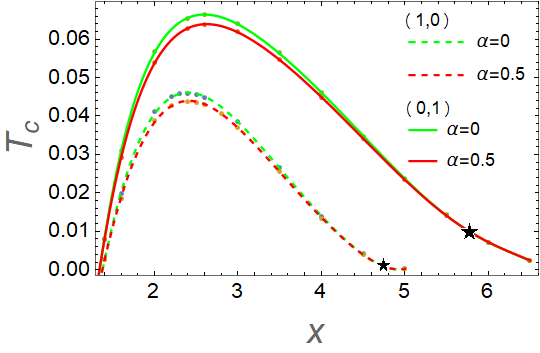}}
 \subfigure[ left endpoint $x_1(\alpha)$ ]
 {\includegraphics[scale=0.26]{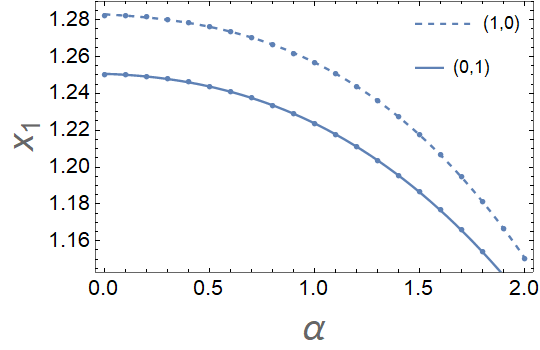}}
 \subfigure[ right endpoint $x_2(\alpha)$ ]
 {\includegraphics[scale=0.25]{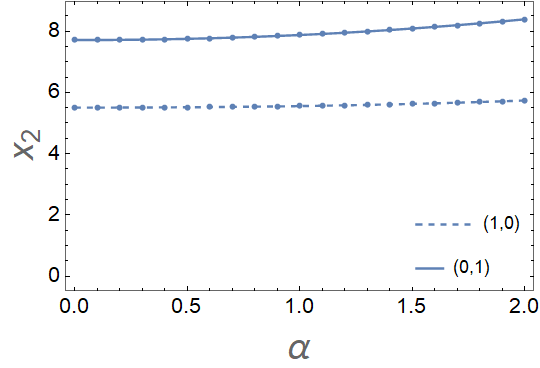}}
  \par\end{centering}
  %\end{multicols}
  \caption{\label{fig:Figure 13}  The roles of $\alpha$.(a)  Changing $\alpha$ doesn't make much difference, but interchanging $q_A$ and $q_B$ makes a big difference.  Two curves for $\alpha=0$ and $\alpha=0.5$ cross at $\star$. (b,c)The left endpoint $x_1$ and the right endpoint $x_2$ of the superconducting region. We fix $a=10, b=4/3, c=14/3$ and choose different $\alpha$. The dashed line represents $(1,0)$, and the solid line represents $(0, 1)$.  }
\end{figure}

In this appendix, we will show the role of $a,b,\alpha$ in Figure \ref{fig:Figure 10} - Figure \ref{fig:Figure 13}. Most of the calculations are unchanged as we have calculated in Section 2. First, we figure out the role of $a$. The left endpoint $x_1$ moves right and the right endpoint $x_2$ moves left when $a$ increases. The optimal temperature $T_O$ increases as $a$ decreases. The superconducting dome expands when $a$ decreases.
Then, we figure out the role of $b$. $x_1$ moves right and $x_2$ moves left to make the dome smaller when $b$ increases. The dependence on $b$ is similar to that on $a$, because $a$ and $b$ are symmetric in our model. As a result, we could conclude that the superconducting dome is expanding as the couplings $Z_A(\chi)=1-\frac{a\,\chi^2}{2}\,,Z_B(\chi)=1-\frac{b\chi^2}{2}\,,Z_{AB}(\chi)=\frac{c\chi^2}{2}$ increase. However, Figure \ref{fig:Figure 13} shows that changing $\alpha$ doesn't make much difference.

Then, we deal with the superconducting dome with the coefficient $\alpha$. We plot the dependence of the boundary of the pseudo-insulating phase ($x_0, T_0$) with the translational symmetry broken by the neutral scalars. As illustrated in Figure \ref{fig:Figure 12}, the pseudo-insulating phase occurs only if $\alpha\geq\alpha_c$ ($\alpha_c\approx0.567358$).
Notice that the superconducting phase with $(q_A,q_B)=(1,0)$ was already studied in \cite{Baggioli:2015dwa}, and our result in this case is consistent with theirs as one can see in Figure \ref{fig:Figure 13}.

\section*{Acknowledgements}
  We would like to thank Xian-Hui Ge, Blaise Gout\'{e}raux and Li Li for valuable comments and discussions. Wenhe Cai is also supported by NSFC China (Grant No. 11805117 and No. 11875184).

  This  work is supported by Mid-career Researcher Program through the National Research Foundation of Korea grant No. NRF-2016R1A2B3007687.

%\bibliographystyle{JHEP}
%\bibliography{Refs_SO.bib}

\end{document}